\documentclass{article}

\usepackage[resetlabels,labeled]{multibib}
\newcites{main,SI}{References, Supporting References}

\usepackage{arxiv}
\usepackage[utf8]{inputenc}
\usepackage{graphicx}
\usepackage{amsmath}

\usepackage[utf8]{inputenc} 
\usepackage[T1]{fontenc}    
\usepackage{hyperref}       
\usepackage{url}            
\usepackage{booktabs}       
\usepackage{amsfonts}       
\usepackage{nicefrac}       
\usepackage{microtype}      
\usepackage{lipsum}
\usepackage{amssymb}

\title{Lifetime-resolved Photon-Correlation Fourier Spectroscopy}

\author{
  Hendrik Utzat\\
  Department of Materials Science and Engineering\\
  Stanford University\\
  Stanford, CA 94305 \\
  \texttt{hutzat@stanford.edu} \\
   \And
 Moungi G. Bawendi \\
  Department of Chemistry\\
  Massachusetts Institute of Technology\\
  Cambridge, MA 02139 \\
  \texttt{mgb@mit.edu} \\
}

\begin{document}
\maketitle

\begin{abstract}
The excited state population of single solid-state emitters is subjected to energy \textit{fluctuations} around the equilibrium driven by the bath and \textit{relaxation} through the emission of phonons or photons. Simultaneous measurement of the associated spectral dynamics requires a technique with a high spectral and temporal resolution with an additionally high temporal dynamic range. We propose a pulsed excitation-laser analog of Photon-Correlation Fourier Spectroscopy (PCFS), which extracts the lineshape and spectral diffusion dynamics along the emission lifetime trajectory of the emitter, effectively discriminating spectral dynamics from \textit{relaxation} and bath \textit{fluctuations}. This lifetime-resolved PCFS correlates photon-pairs at the output arm of a Michelson interferometer in both their time-delay between laser-excitation and photon-detection $T$ and the time-delay between two photons $\tau$. We propose the utility of the technique for systems with changing relative contributions to the emission from multiple states, for example, quantum emitters exhibiting phonon-mediated exchange between different fine-structure states.
\end{abstract}

\section{Introduction}
\begin{figure}[htp]
    \centering
    \includegraphics[width=12cm]{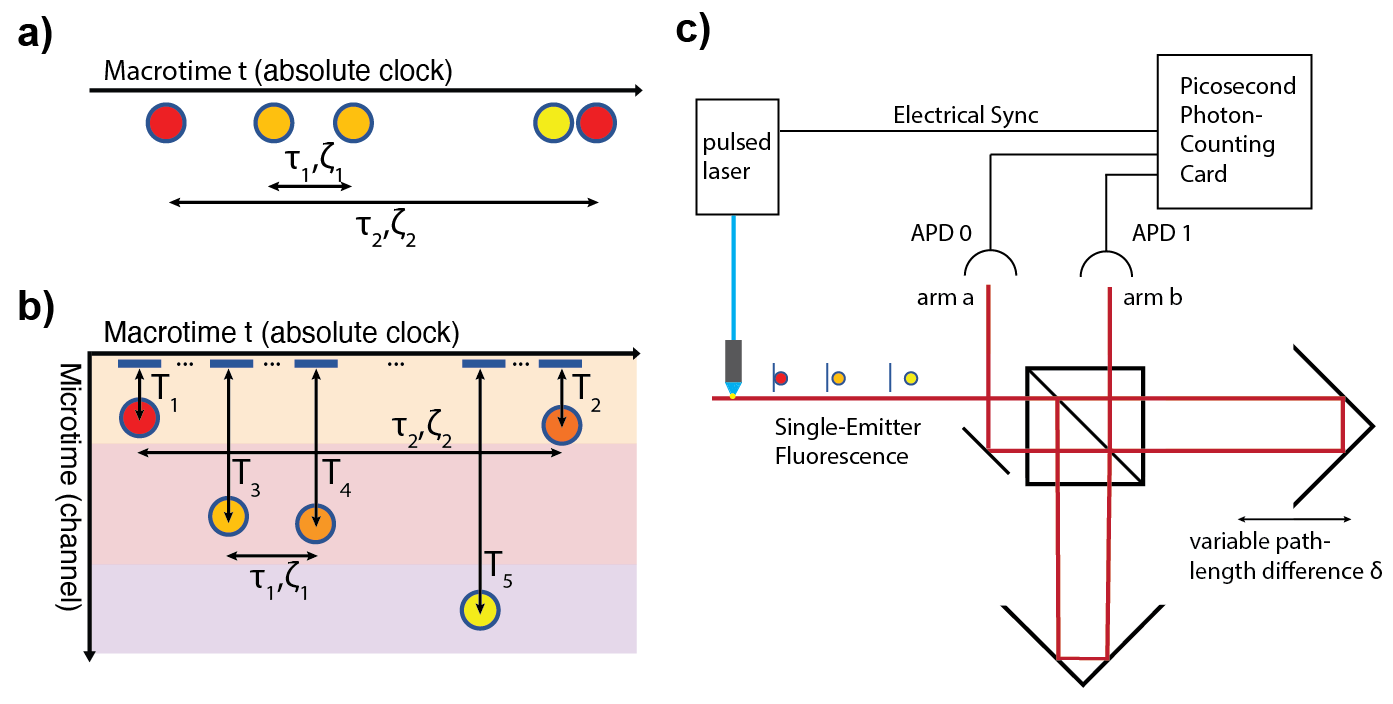}
    \caption{In conventional PCFS, the spectral correlation is compiled from photon-pairs irrespective of their microtime T, often under continuous wave excitation (a). In lifetime-resolved PCFS, photon-pairs with a given microtime $T$ and macrotime separation $\tau$ are spectrally correlated (b). Here, we adopt a time-binning approach to collect photons with different T in suitable microtime intervals as indicated by the color-shaded background. The proposed optical setup is shown in (c). The photon-stream from a single emitter under pulsed excitation is directed into a variable path-length difference Michelson interferometer. All photon-counts at the output arms of the interferometer are recorded in time-tagged (T3) mode using picosecond single-photon counting electronics.}
    \label{fig:Fig1}
\end{figure}

The spectral dynamics of single emitters can broadly be categorized as fluctuations and relaxation. Fluctuations are temporally stochastic variations around the equilibrium configuration of any chemical system interacting with its environment at non-zero temperature. Relaxation refers to the system's return to the ground state equilibrium after preparation of a non-equilibrium state, for example, through a laser-driven creation of an excited state population. For single optical emitters, fluctuations can manifest as spectral diffusion, which is spectral jumping occurring from nanoseconds to seconds that is reflective of the microscopic interaction of the bath with the excited state.\cite{Kettner1994,Ambrose1991} Relaxation occurs via irreversible phonon- or spin-mediated dissipation or spontaneous emission of photons, processes typically observed from picoseconds to microseconds.\cite{Masia2012}\\
Simultaneous measurement of the full relaxation and fluctuation dynamics for single emitters requires a technique with high spectral and temporal resolution with an additionally high temporal dynamic range from picoseconds to seconds. No single such technique exists and different approaches present their own strengths and weaknesses. Streak-cameras can resolve the spectral evolution along the photoluminescence lifetime trajectory with picosecond resolution, but fail to resolve any dynamics beyond a few nanoseconds and typically lack single-molecule sensitivity.\cite{ReschGegner2008} CCD-based single-molecule emission spectroscopy can have millisecond temporal resolution, especially if auto-correlation of individual spectra recorded at high sweep-rates is performed, but fails to resolve even faster fluctuations and does not discriminate spectral dynamics along the lifetime trajectory.\cite{Plakhotnik1998,Plakhotnik1999}
Hong-Ou-Mandel (HOM) spectroscopy has been used to measure the coherence of single photons through two-photon interference. HOM can resolve energy fluctuations through a decrease in photon-coalescence efficiency on fast (picosecond to nanosecond) timescales and lifetime-resolved photon-presorting is straightforward.\cite{Kuhlmann2015, Thoma2016} However, HOM spectroscopy is exclusively suitable for single-photon emitters at low temperatures as it requires photon-coherences near the transform limit.
Moreover, the dynamic range of HOM is practically limited to a few nanoseconds delay-time between the two interfering photons, rendering HOM of limited utility for measuring fluctuations occurring over many orders of magnitude in time. Cross-correlating spectrally-filtered photons provides higher temporal dynamic range, but is limited by the finite bandwidth of optical filters restricting the technique to broad lineshapes and spectral diffusion with large energetic spread.\cite{Sallen2010}\\ Fourier spectroscopy is not bound by limitations in spectral bandwidth or two-photon delay-times as each photon self-interferes. As a result, Fourier spectroscopy can readily be married with photon correlation to provide spectral readout with high temporal dynamic range and with arbitrarily high temporal resolution only limited by photon shot-noise.\cite{Brokmann2006} This technique, Photon Correlation Fourier Spectroscopy (PCFS), has now been established as a powerful tool for the study of optical dephasing and spectral fluctuations of single emitters at low and room temperatures.\cite{Coolen2008a,Utzat2019,Cui2013a}Despite the success in characterizing spectral fluctuations, so far PCFS was not able to resolve any spectral changes associated with relaxation of the system back to the ground state, for example phonon-mediated relaxation or energy transfer between different states.\\
Here, we propose a pulsed excitation-laser analog of PCFS that readily extracts relaxation and fluctuation dynamics from single emitters. The proposed technique is shown in Figure \ref{fig:Fig1}. In conventional PCFS, all photons emitted after the continuous-wave laser excitation are used to compile the spectral correlation (a), the auto-correlation of the spectrum compiled from photon pairs with a temporal separation of $\tau$ along the macrotime axis $t$ of the experiment and an energy difference of $\zeta$. In lifetime-resolved PCFS, photon-pairs are additionally correlated in the microtime $T$ after pulsed laser excitation. Specifically, the photons are binned according to their microtime, sometimes referred to as the TCSPC channel, and spectral correlations are calculated using these microtime-separated photons (b).
The technique can be readily implemented using picosecond single-photon counting equipment as shown in Fig. 1c and requires high-throughput post-processing photon-correlation analysis. We show through numerical simulation that this lifetime-resolved PCFS technique can separate the lineshapes and spectral diffusion dynamics of systems with more than one emissive state as long as the relative weights of the emission from different states changes over the course of the photoluminescence lifetime. More broadly, lifetime-resolved PCFS can in principle extract spectral fluctuations and relaxation with temporal resolutions only limited by the IRF response time of single-photon counting modules. 

\section{Theoretical Derivation and Numerical Simulation}\label{theory}
\subsection{Lifetime-resolved PCFS}
In PCFS, the interferometer path-length difference is adjusted to discrete positions inside the coherence length of the emission and periodically dithered on second timescales and over a multiple of the emitter center wavelength.\cite{Brokmann2006} The dither introduces anti-correlations in the intensity cross-correlation functions of the output arms that encode the degree of spectral coherence at a given center position. The lineshape dynamics is thus encoded in the intensity correlations as a function of the time-separation between photons $\tau$. We show in the Supplementary Information that the PCFS equations can straightforwardly be expanded to include spectral dynamics along the microtime $T$. The central observable in lifetime-resolved PCFS for a spectrum $s(\omega,t,T)$ dependent on the microtime $t$ and macrotime $T$ is given by the spectral correlation $p(\zeta,\tau,T)$ as

\begin{equation}\label{spec:corr_central1}
p(\zeta,\tau,T)=\langle\int_{-\infty}^{\infty}s(\omega,t,T)*s(\omega+\zeta,t+\tau,T) d\omega\rangle, \end{equation}
where $\langle...\rangle$ represents the time average. Equation \ref{spec:corr_central1} describes the central observable in lifetime-resolved PCFS and can be understood intuitively as a histogram of photon-pairs with a shared microtime $T$, a macrotime separation of $\tau$, and an energy separation $\zeta$. The form and interpretation of $p(\zeta,\tau,T)$ depend on the dynamics of the emissive system and will be discussed in the following sections.\\
We first discuss the general form of the spectral correlation for a system undergoing spectral diffusion in section \ref{Spec_diff}. We then discuss two universal systems that map onto many specific real-world scenarios. First, we consider a system of two uncoupled and lifetime-distinct radiating dipoles undergoing uncorrelated spectral diffusion in section \ref{static doublet_Gauss}. Second, we consider a system of two coupled radiating dipoles subject to population exchange and correlated spectral diffusion in section \ref{static doublet_Gauss_and_relax}.

\subsection{The effect of spectral fluctuations on the spectral correlation}\label{Spec_diff}
We consider equation \ref{spec:corr_central} for spectral fluctuations $\delta\omega(t,T)$ that present along the macrotime axis of the experiment around the center frequency $\omega_{0}$ of a spectrum. We can write for the spectrum $s(\omega,t,T)=s(\omega,T) \otimes \delta(\omega-\delta\omega(t,T))$, where $\otimes$ is the convolution, $s(\omega,T)$ the undiffused spectrum, and $\delta\omega(t,T)$ the time-dependent shift from the center wavelength. Spectral fluctuations can be characterized by the correlation function $C(\tau)=\langle\delta\omega(t,T)\delta\omega(t+\tau, T)\rangle$. The canonical form of any spectral correlation can then be recast as

\begin{equation}\label{spec_corr_fluc}
\begin{split}
p(\zeta,\tau,T)=\langle\int_{-\infty}^{\infty} s(\omega,t,T) s(\omega +\zeta,t+\tau,T)  d\omega\rangle\\
=C(\tau)p(\zeta,\tau \rightarrow 0,T)+[1-C(\tau)]p(\zeta,\tau \rightarrow \infty,T),
\end{split}
\end{equation}

reflecting the transition from the undiffused spectral correlation (absent any fluctuations $p(\zeta,\tau \rightarrow 0, T)$) to the diffused spectral correlation $p(\zeta,\tau \rightarrow \infty, T)$ with the evolution of $C(\tau)$. Note that for $\tau \rightarrow 0$, $\delta\omega(t_{1},T)=\delta\omega(t_{2},T)$ and the spectral correlation thus reduces to the homogeneous spectral correlation $p(\zeta,\tau\rightarrow 0,T)=\langle\int_{-\infty}^{\infty}s(\omega,T)s(\omega-\zeta,T)d\omega\rangle$.

\subsection{A lifetime-distinct doublet undergoing Gaussian spectral fluctuations}\label{static doublet_Gauss}
\begin{figure}[htp]
    \centering
    \includegraphics[width=12cm]{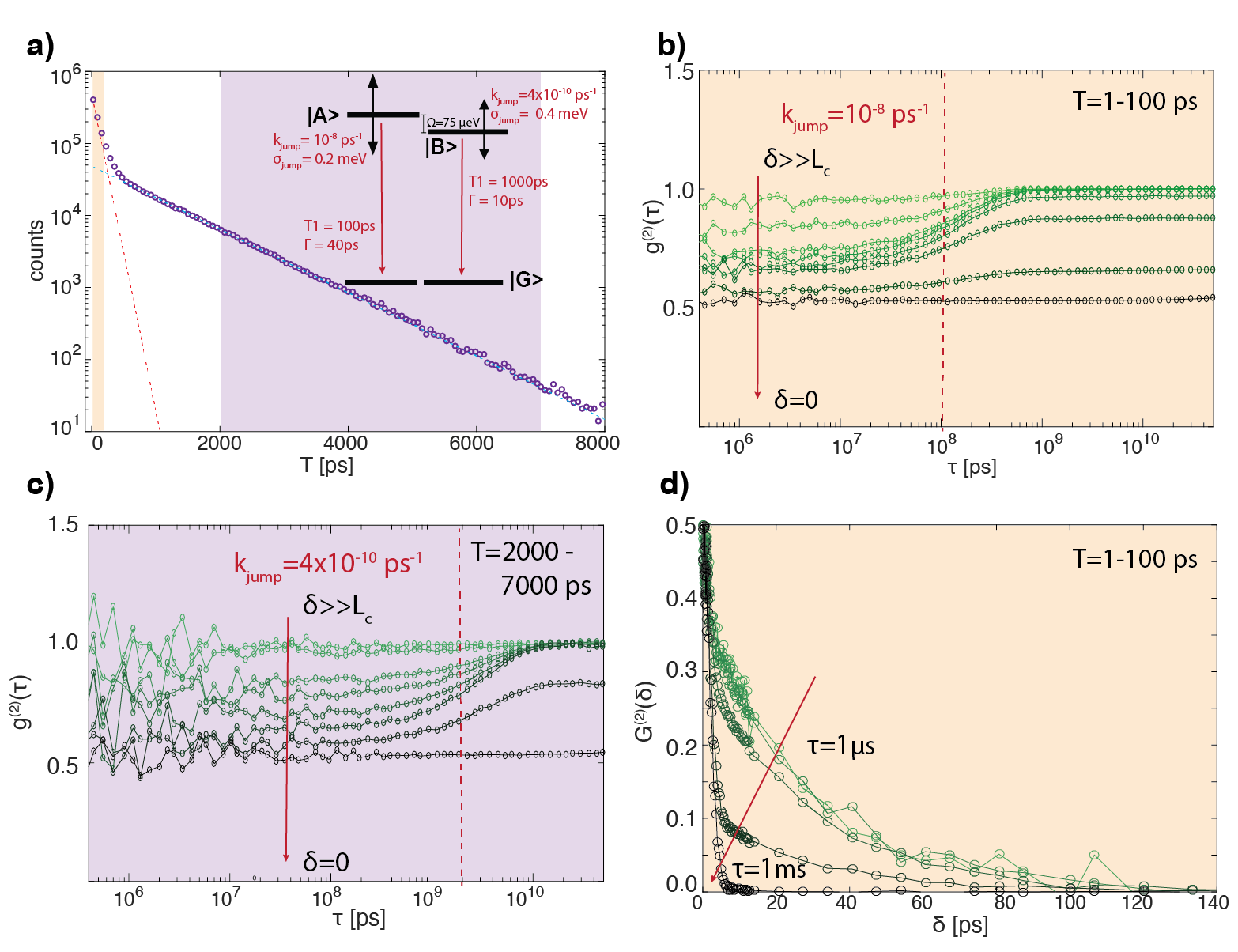}
    \caption{Simulation of two uncoupled radiating dipoles involving states $|A\rangle$ and $|B\rangle$. The two transitions are coupled to two different bath fluctuations and exhibit different lifetimes $T_{1}$, linewidths $\Gamma$ and spectral diffusion parameters $k_{jump}$ and $\sigma_{A,B}$. The total fluorescence lifetime of the system exhibits a biexponential decay (a). The shaded panels (b) and (c) show the cross-correlation functions $g_{X}^{(2)}(\tau)$ for different optical path-length differences $\delta_{0}$ and microtimes $T$, where $|A\rangle$ and $|B\rangle$ are the dominant emissive states, respectively. The loss of coherence with increasing $\tau$ is evident from the reduction in anti-correlation. This coherence loss occurs at earlier $\tau$ for early-$T$ photons (emission predominantly from $|A\rangle$, (b)) compared to late-$T$ photons (emission predominantly from $|B\rangle$, (c)). The PCFS interferogram $G^{(2)}(\delta,\tau)$ for early-T photons is shown in (d) and reflects the evolution from the exponential homogeneous dephasing at early $\tau$ to the spectrally-diffused Gaussian dephasing at late $\tau$.}
    \label{fig:Fig3}
\end{figure}

\begin{figure}[htp]
    \centering
    \includegraphics[width=12cm]{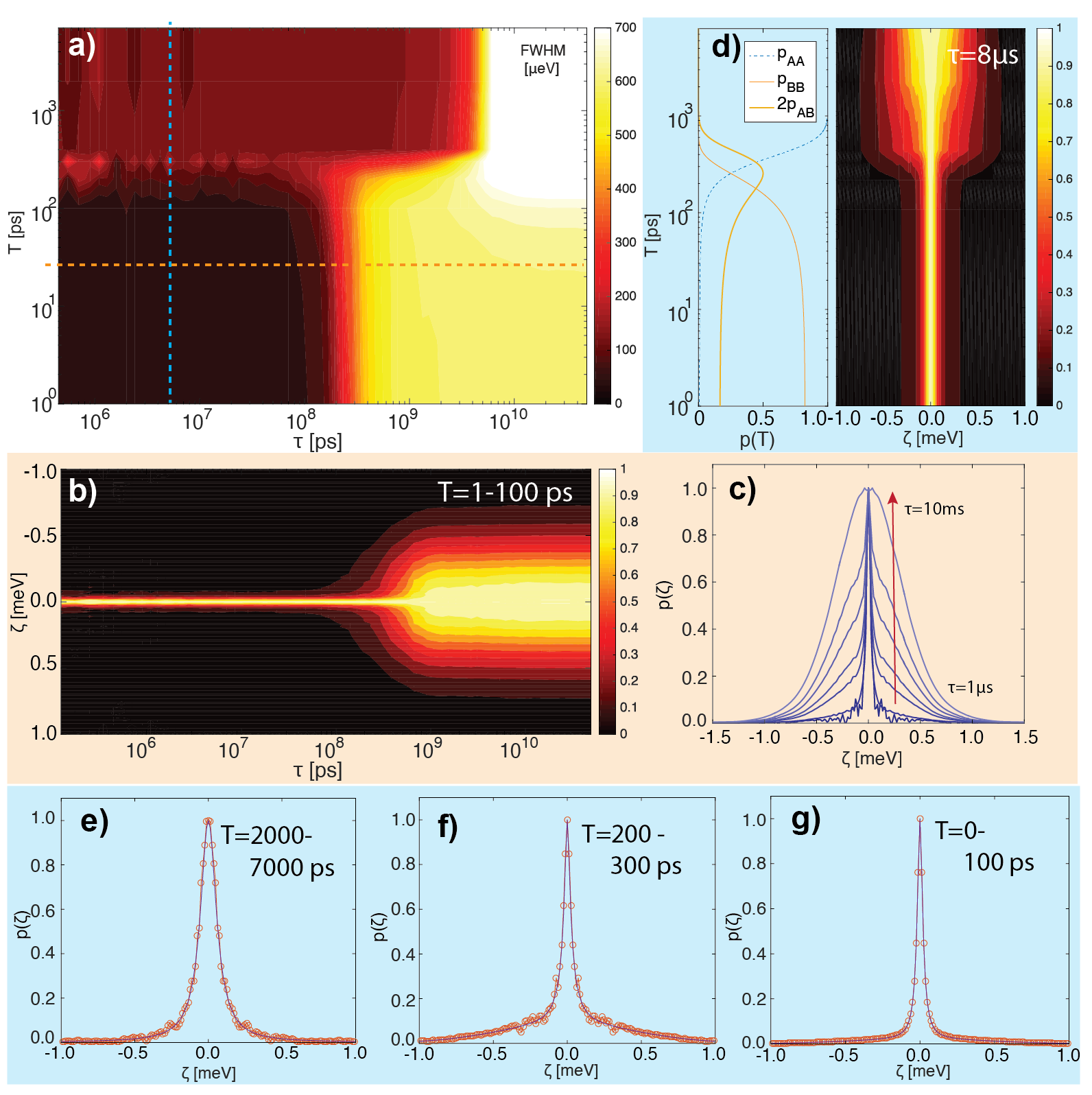}
    \caption{Spectral results of the lifetime-resolved PCFS simulation of two uncoupled dipoles. (a) shows the full-width-at-half-maximum (FWHM) of $p(\zeta,\tau,T)$ along $T$ and $\tau$. The difference in the homogeneous linewidths of $|A\rangle$ and $|B\rangle$ at early $\tau$ and in the diffused linewidths at late $\tau$ are immediately obvious in this representation. The orange-shaded panels (b) and (c) show the effect of spectral diffusion for early-microtime photons originating mostly from state $|A\rangle$. We show the evolution of the weights of auto- and cross-correlations between states along $T$ in (d).The weights are derived from the relative amplitude of the two exponential components of the photoluminescence decay in (Fig.\ref{fig:Fig3}a). Taking $p_{AA}$,$p_{BB}$, and $p_{AB,BA}$ into account, we apply a global fit to the spectral correlation along $T$  to recover the lineshape parameters of the undiffused system as shown in (e),(f) and (g). The broad underlying Gaussian component in (f) reflects the cross-correlation of the diffused distributions of $|A\rangle$ and $|B\rangle$ and has a width of $\sigma\approx\sqrt{\sigma_{A}^2+\sigma_{B}^2}$.}
    \label{fig:Fig4}
\end{figure}

We discuss a system of two uncoupled and lifetime-distinct radiating dipoles undergoing uncorrelated spectral diffusion and involving states $|A\rangle$ and $|B\rangle$. The system's energy diagram is shown in figure \ref{fig:Fig3}a (inset). The microscopic interpretation involves a system with two emissive states coupled to different bath fluctuations. We show how lifetime-resolved PCFS can separate the homogeneous lineshape and spectral diffusion parameters of the the two transitions.\\
The different emission lifetimes result in microtime-dependent relative weights of emission intensity originating from states $|A\rangle$ and $|B\rangle$ after equal populations have been prepared through laser excitation. We decompose the overall dynamic spectrum of the system $s(\omega,t,T)$ into microtime-dependent components as $s(\omega,t,T)=a(T)s_{A}(\omega,t)+b(T)s_{B}(\omega,t)$, where $a(T)$ and $b(T)$ are the relative probabilities of a given photon originating from either state $|A\rangle$ or $|B\rangle$, and show that the spectral correlation expands as
\begin{equation}\label{reduced:spec_corr}
\begin{split}
p(\zeta,\tau,T)=a(T)^{2}p_{AA}(\zeta,\tau)\\
+ a(T)b(T)(p_{AB}(\zeta,\tau)+p_{BA}(\zeta,\tau))\\
+ b(T)^{2}p_{BB}(\zeta,\tau)
\end{split}
\end{equation}
(see Supplementary Information). The terms quadratic in $a(T)$ and $b(T)$ represent the spectral auto-correlations of the individual states $p_{AA}$ and $p_{BB}$, while the cross-terms involving $p_{AB}$ represent the cross-correlation of the spectra $s_{A}(\omega,t,T)$ and $s_{B}(\omega,t,T)$. The form of the spectral correlation can be understood intuitively because the spectral correlation is compiled from pairs of photons with origins drawn from the four possible combinations of $|A\rangle$ and $|B\rangle$. Importantly, the left- and right-sided correlations $p_{AB}$ and $p_{BA}$ are not identical unless $s_{A}(\omega)$ and $s_{B}(\omega)$ share the same center frequency $\omega_{0}$ and are symmetric in $\omega$.\\
Spectral diffusion is a ubiquitous process observed for many single emitters. Common descriptions of single-emitter spectral diffusion are the non-Markovian and discrete Poissonian Wiener process\cite{Beyler2013} or the mean-reverting Ornstein-Uhlenbeck process.\cite{Richert2001} These processes describe spectral diffusion phenomenologically and for simplicity we consider a simple non-Markovian Poissonian Gaussian jumping model (GJM).\cite{Utzat2019} The GJM process is characterized by a time-invariant probability density for discrete spectral jump occurrence to a new spectral position drawn from a Gaussian probability distribution function over $\omega$. For the two states $|A\rangle$ and $|B\rangle$ as denoted in the subscripts, we write  $Prob(\delta \omega_{A,B})=\frac{1}{\sigma_{A,B}\sqrt{2\pi}}e^{-\frac{\delta\omega^{2}}{2\sigma_{A,B}^{2}}}$ for the probability of a given spectral shift at a point in time. Here, we have introduced the spectral fluctuation term $\delta\omega_{A,B}$ from earlier. The microscopic interpretation of this process is the time-stochastic variation of the bath assuming discrete conformations coupling to the system. The corresponding fluctuation correlation function can be written as $C(\tau)=e^{-\tau/\tau_{c}}$ and is described by an exponential decay with a characteristic spectral jump time of $\tau_{c}$. When the two states diffuse independently of each other, no correlation is present and $C_{AB}(\tau)=0$. In this case, $\langle\delta\omega_{A}(t)\delta\omega_{B}(t+\tau)\rangle=\langle\delta\omega_{a}(t)\rangle\langle\delta\omega_{B}(t+\tau)\rangle=0$ because independently diffusing emissive states will not be correlated and the cross-terms $p_{AB}$ and $p_{BA}$ in equation \ref{reduced:spec_corr} only reflect the cross-correlations of the inhomogeneous components $p_{AB/BA}(\zeta,\tau \rightarrow \infty)$. Absent any memory of spectral fluctuations even at early $\tau$, the time average over the spectral-correlations of all random configurations is the cross-correlation of the inhomogeneously broadened (diffused) spectra

\begin{equation}
p_{AB}(\zeta)=\langle\int_{-\infty}^{\infty}e^{-\frac{\delta\omega^{2}}{2\sigma_{A}^{2}}}e^{-\frac{(\delta\omega+\zeta)^{2}}{2\sigma_{B}^{2}}}d\delta\omega\rangle,
\end{equation}\label{equation:idk}

where $\sigma_{A}$ and $\sigma_{B}$ are the widths of the Gaussian probability envelopes of the diffused distributions of states $|A\rangle$ and $|B\rangle$.\\
We numerically simulate the system of independently-diffusing optical transitions with parameters commensurate with typical experimental cryogenic single-molecule spectroscopy (see Supplementary Information). The time-domain results of the simulation are discussed in Figure \ref{fig:Fig3}. The configuration of the system is shown in Figure \ref{fig:Fig3}(a). The corresponding lifetime exhibits biexponential decay behavior as expected. In (b) and (c), we compare the cross-correlation functions for two different slices with microtime ranges of $T=0-100$ps and $T=2000-7000$ps, where $|A\rangle$ and $|B\rangle$ are the dominant emissive states, respectively. Unlike for the static doublet discussed in the Supplementary Information, the cross-correlations $g_{X}^{(2)}(\tau)$ indicate spectral dynamics evident from the loss of anti-correlation at longer $\tau$. As we specify different jumping rates for the two states, the decay of the spectral coherence evident in (b) and (c) occurs at different $\tau$. The PCFS interferogram derived from the cross-correlations (see Supplementary Information for the derivation) for photons emitted with a time constant of $<100ps$ is shown in (d) and informs on the loss of photon-coherence between $1\mu\textrm{s}$ and $1\textrm{ms}$ owing to the energy fluctuations of the photons emitted $<100$ ps after laser excitation.\\
In Figure \ref{fig:Fig4} we discuss the same simulation results in the spectral domain. In (a), we show the full-width-at-half-maximum (FWHM) of the spectral correlation for both $T$ and $\tau$, a representation that makes immediately obvious the differences in the homogeneous linewidths at early $\tau$ and the differences in spectrally-diffused linewidths at late $\tau$. For completeness we also show $p(\zeta,T)$ (d) and $p(\zeta,\tau)$ (b) for fixed $\tau$ and $T$, respectively. These two representations inform on the spectral evolution owing to spectral diffusion and changing relative emission contributions from different states, respectively. (c) displays the evolution of the spectral correlation from the narrow homogeneous spectrum with a Lorentzian lineshape to the diffused Gaussian lineshape.\\
One capability of lifetime-resolved PCFS is the ability to extract the homogeneous linewidths of different lifetime-distinct states in the presence of fast spectral diffusion. We demonstrate this ability through a global fit to the T-dependent spectral correlation. We define a model for the fit as a linear combination of two Lorentzians and a Gaussian with floating linewidths parameters. The relative amplitudes $p_{AA}$,$p_{BB}$, and $p_{AB,BA}$ are calculated according to \ref{reduced:spec_corr} taking the weights $a(T)$ and $b(T)$ from fits to the emission lifetime into account. $p_{AA}$,$p_{BB}$, and $p_{AB,BA}$ are also displayed in (d). We apply a global fit to the slices of the spectral correlation $p(\zeta,\tau=60 \mu \textrm{s}, T)$ along $T$ as shown in (e),(f) and (g). The cross-correlation $p_{AB,BA}$ present as a broad Gaussian background superimposed with the homogeneous Lorentzian spectral correlations $p_{AA,BB}$ as introduced in equation \ref{reduced:spec_corr}. The width of this Gaussian component is $\sigma_{AB}\approx\sqrt{\sigma_{A}^{2}+\sigma_{B}^{2}}$. The homogeneous lineshape parameters parsed into the numerical model are extracted by the fit within photon shot-noise, thus validating the approach adapted herein.
We note that in PCFS, the high temporal resolution achieved through photon-correlation comes at the cost of the loss of the absolute phase of the spectral information. In other words, both the asymmetry of the lineshape and the center frequency of $s(\omega)$ is lost in the spectral correlation $p(\zeta)$. The unambiguous reconstruction of $s(\omega)$ from $p(\zeta)$ is therefore impossible and the spectral correlation is typically fit with a model parametrizing a suitable form for the underlying emission spectrum, as we adapted herein. \cite{Cui2013a,Utzat2019}

\subsection{A dynamic doublet with population transfer and spectral fluctuations}\label{static doublet_Gauss_and_relax}
\begin{figure}[htp]
    \centering
    \includegraphics[width=12cm]{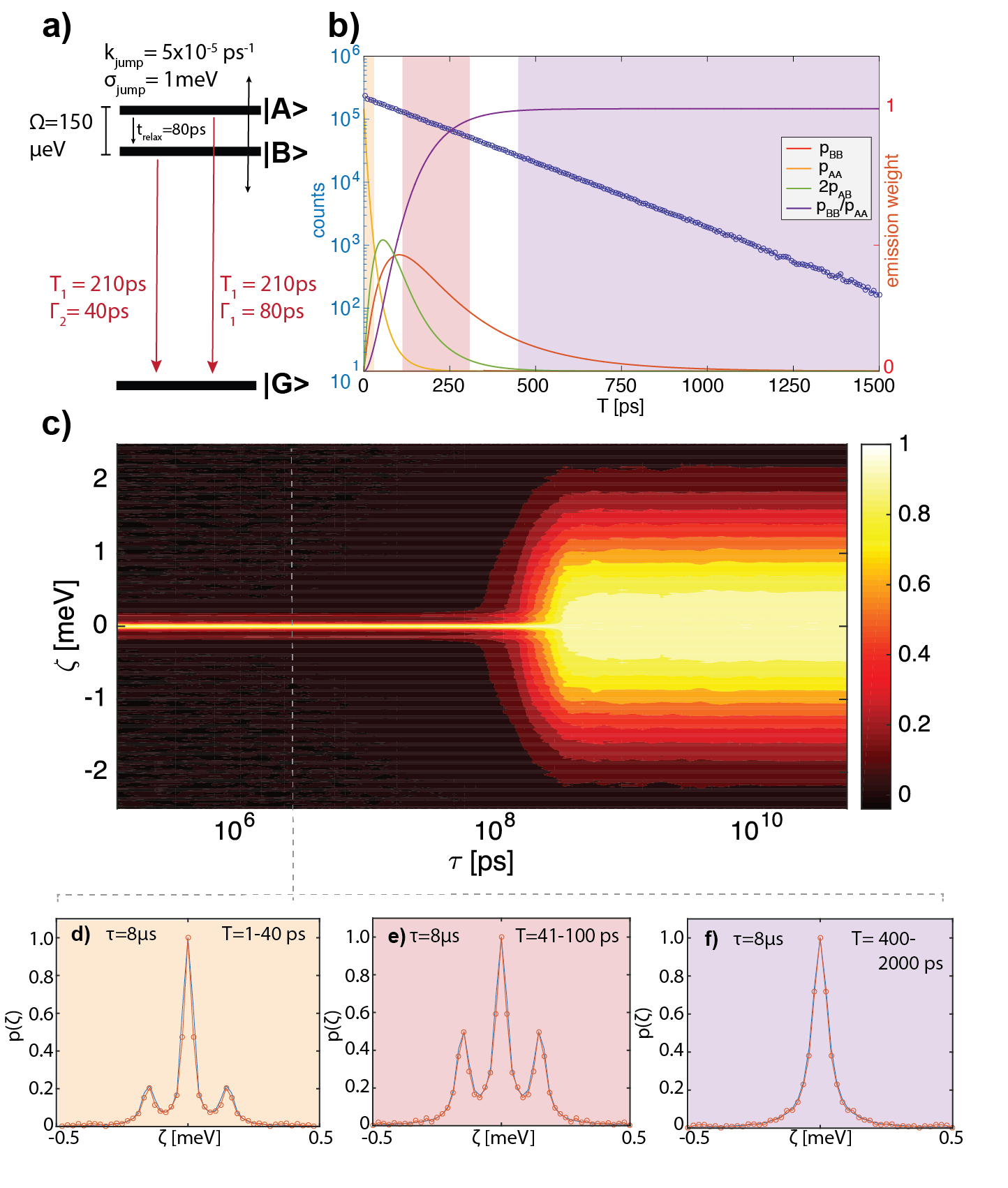}
    \caption{Lifetime-resolved PCFS simulation of two coupled dipoles undergoing population transfer and interacting with the same bath resulting in collective spectral diffusion of the doublet (a). We introduce a phonon-mediated relaxation rate between the upper and lower state of $k_{relax}=1/80 \textrm{ps}^{-1}$. As the radiative rates of the two states are chosen to be equal, the emission lifetime follows a monoexponential decay behavior despite changing relative populations of $|A\rangle$ and $|B\rangle$ with the microtime (b). The spectral correlation irrespective for all photons irrespective of their microtime is shown in (c) and demonstrates the transition from a triplet at early $\tau$ to the spectrally-diffused distribution at late $\tau$. The fine-structure splitting $\Omega$, the linewidths $\Gamma_{A,B}$, and the relaxation rate $k$ can be recovered through lifetime-resolved PCFS and a global fit of the slices along $T$ with a fixed macrotime correlation of $\tau=8 \mu s$ (d),(e) and (f).}
    \label{fig:Fig5}
\end{figure}

We now turn to a system of two coupled radiating dipoles undergoing population exchange and subject to correlated spectral diffusion. A specific example would be solid-state quantum emitters undergoing incoherent and phonon-mediated population transfer after non-resonant excitation. \cite{Vinattieri1994} In quantum emitters, disentangling the relaxation rate and coherence times of the different fine-structure states in the presence of spectral diffusion is important for a detailed understanding of the dephasing process as phonon-mediated population exchange constitutes an important dephasing process in the solid-state.\cite{Masia2012} We depict the system's energy diagram in Figure \ref{fig:Fig5}(a), which exhibits two excited states with equal oscillator strengths and an irreversible relaxation rate $k$ from the higher to the lower-lying state. In this system, photon emission from the higher-lying state $|A\rangle$ will start immediately after population of the state. Emission of the lower-lying state $|B\rangle$ requires further relaxation and is often phonon-mediated.\cite{Masia2012} The relative population of states $|A\rangle$ and $|B\rangle$ will thus change during the emission lifetime of the overall system as long as the relaxation rate $k$ is faster than the radiative rate $1/T_{1}$ of both $|A\rangle$ an $|B\rangle$. The population dynamics of the system can be described by the following set of coupled equations:

\begin{equation}
\frac{d|A\rangle}{dt}=-(k+1/T_{1})|A\rangle
\end{equation}

\begin{equation}
\frac{d|B\rangle}{dt}=k|A\rangle-1/T_{1}|B\rangle
\end{equation}
with the solutions:

\begin{equation}
|A\rangle(t)=|A\rangle_{0}e^{-(k+1/T_{1})t}
\end{equation}

\begin{equation}
|B\rangle(t)=-e^{-(1/T_{1}+k)t}+Ce^{-1/T_{1}t}.
\end{equation}

We show the effect of the changing relative cross-correlation probabilities between states $|A\rangle$ and $|B\rangle$ ($p_{AA}$,$p_{BB}$) in Figure \ref{fig:Fig5} (b). Despite the $T$-invariant exponential population decay constant leading to a monoexponential photoluminescence lifetime of the overall system, the relative weights of $p_{AA}$ and $p_{BB}$ are changing with $T$.We show the spectral correlation of the lifetime-resolved PCFS experiment with indiscriminate $T$ in (c). On timescales shorter than the spectral diffusion time $\tau$, the fine-structure states are well-separated. At late $\tau$, the broad diffused lineshape obfuscates the fine-structure splitting.\\
We demonstrate that lifetime-resolved PCFS can recover the lineshape parameters of the homogeneous doublet by applying a least-squares fit of a suitable model to the T-dependent spectral correlation as shown in (d),(e),(f). The model consists of two Lorentzians. with the floating linewidths $\Gamma_{1}$, $\Gamma_{2}$, an energy offset $\Omega$ and a relaxation rate $k$, which determines the temporal change of the relative emission contributions of $\langle A \rangle$ and $\langle B \rangle$. We recover all model parameters within photon shot-noise thus validating the utility of lifetime-resolved PCFS to extract the coherences and relaxation rates of different emissive fine-structure states. We note that the observation of early-$\tau$ multiplets in the spectral correlation compared to the broad Gaussian background in section \ref{static doublet_Gauss} is the signature of correlated spectral diffusion dynamics between the two states. Our simulations suggest that measuring the photon-coherences of quantum emitters exhibiting spectral fluctuations and different emissive fine-structure states will provide an avenue to study quantum emitter optical dephasing through both fluctuations and population exchange between different electronic states. 

\section{Conclusions}
We propose a new photon-correlation spectroscopic technique that extracts spectral fluctuations along the lifetime-trajectory of single emitters. The technique works through time-correlation of photons detected at the output arms of a variable path-length difference interferometer in both the microtime and macrotime domain and can be implemented using standard picosecond photon-counting electronics. We show that lineshape and fluctuation parameters can be extracted from the fits to the lifetime-resolved spectral correlations. Our technique opens up multiple frontiers in single-emitter spectroscopy. We emphasize that our technique is general, but point to its special utility in quantum emitter research enabled by the high spectral resolution required to resolve photon-coherences at low temperatures. Experimental efforts will be directed towards probing the fluctuation dynamics of non-stationary systems and investigation of the decoherence processes in quantum emitters. Specific materials are readily available such as emissive defects in diamond and emerging 2D materials as well as semiconductor nanostructures.

\section{Acknowledgements}
The lead author of this study (H.U., study conception, derivation, modeling and interpretation) was initially funded by the U.S. Department of Energy, Office of Basic Energy Sciences, Division of Materials Sciences and Engineering (award no. DE-FG02-07ER46454) and funded by Samsung Inc. (SAIT) during the completion of the study. We thank Weiwei Sun, David Berkinsky, Alex Kaplan, Andrew Proppe, and Matthias Ginterseder for critically reading the manuscript and their feedback.


\newpage

\newpage

\newpage

\newpage

\newpage

\bibliographystyle{ieeetr}
\bibliography{references}

\newpage
\title{Supplementary Information: Lifetime-resolved Photon-Correlation Fourier Spectroscopy}

\setcounter{page}{1}

\huge
Supplementary Information: Lifetime-Resolved Photon-Correlation Fourier Spectroscopy
\normalsize

\section*{Derivation of Lifetime-Resolved PCFS}
 We derive PCFS classically, that is using the field rather than the photon description of light. In practical PCFS experiments, the finite integration time renders the photon anti-bunching timescales outside of reasonable signal-to-noise regimes, justifying our classical approach adapted herein.\citeSI{Brokmann2006A,Utzat2019A} We denote the microtime as $T$, the macrotime of the absolute experimental clock as $t$, and the time-separation between photons as $\tau$. We show how the experiment can access the spectral correlation $p(\zeta,\tau,T)$, a histogram of photon pairs with a temporal separation $\tau$ along the macrotime axis, a separation from the laser excitation pulse $T$, and an energy separation of $\zeta$.\\
The intensity distribution at the two interferometer outputs $a$ and $b$ for an arbitrary spectrum $s(\omega,t,T)$ evolving over the microtime $T$ after laser excitation and the macrotime $t$ of the absolute experimental clock at a given center path-length difference $\delta_{0}$ is given by

\begin{equation}\label{equ:one}
I_{a,b}(\delta_{0},t,T)=\frac{1}{2}I(t)[1\pm FT\{s(\omega,t,T)\}_{\delta(t)}],
\end{equation}

where $\delta(t)$ represents a time-dependent path-length difference defined by the chosen scanning trajectory of the interferometer and exhibiting the form $\delta(t)=\delta_{0}+\Delta\delta(t)$, where $\delta_{0}$ is the center path-length difference and $\Delta\delta(t)$ is a time-dependent term introduced experimentally through periodic dithering of one of the interferometer arms. $FT\{s(\omega,t,T)\}_{\delta(t)}$ is the Fourier transform of the time-dependent single-emitter spectrum. Substitution of \ref{equ:one} into the canonical form for the intensity cross-correlation function $g_{\times}^{(2)}(\tau)$

\begin{equation}\label{equ:two}
g_{\times}^{(2)}(\tau)=\frac{\langle I_{a}(t)I_{b}(t+\tau)\rangle}{\langle I_{a}(t)\rangle \langle I_{b}(t+\tau)\rangle},
\end{equation}
where $\langle...\rangle$ denotes the time average and $I_{a,b}(t,T)$ are the time-dependent intensities at the two detector outputs $a$ and $b$, gives

\begin{equation}\label{equ:four}
g_{\times}^{(2)}(\delta_{0},\tau,T)=\frac{\langle I(t,T)I(t+\tau,T)[1+FT\{s(\omega,t,T)\}_{\delta(t)}][1-FT\{s(\omega,t+\tau,T)\}_{\delta(t+\tau)}]\rangle}{\langle I(t,T)[1+FT\{s(\omega,t,T)\}_{\delta(t)}]\rangle\langle I(t+\tau,T)[1-FT\{s(\omega,t+\tau,T)\}_{\delta(t+\tau)]}\rangle}.
\end{equation}

The exact functional form of $\Delta\delta(t)$ can be chosen arbitrarily as long as it is experimentally ensured that the path-length difference for each recorded correlation function is scanned symmetrically around the center position $\delta_{0}$: $\int_{0}^{T_{ac}}\Delta\delta(t)dt=0$, where $T_{ac}$ is the data acquisition time at each $\delta_{0}$. Under this condition, terms linear in the Fourier transforms of the form $\langle FT\{s(\omega,t,T)\}_{\delta(t)}\rangle$  or $\langle FT\{s(\omega,t+\tau,T)\}_{\delta(t+\tau)}\rangle$ vanish and the expansion of equation \ref{equ:four} can be recast as

\begin{equation}
g_{\times}^{(2)}(\delta_{0},\tau,T)=\frac{\langle I(t,T)I(t+\tau,T)\rangle}{\langle I(t,T)\rangle\langle I(t+\tau,T)\rangle}(1-\langle FT\{s(\omega,t,T)\}_{\delta(t)}\times FT\{s(\omega,t+\tau,T)\}_{\delta(t+\tau)}\rangle).
\end{equation}

The first term is just the intensity auto-correlation function $g_{||}^{(2)}(\tau, T)$.
We then introduce the PCFS interferogram $G^{(2)}(\delta_{0},\tau,T)$ defined as:

\begin{equation}
\label{eqn:five}
G^{(2)}(\delta_{0},\tau,T)=1-\frac{g_{\times}^{(2)}(\delta_{o},\tau, T)}{g_{||}^{(2)}(\delta_{0},\tau, T)}=\langle FT\{s(\omega,t,T)\}_{\delta(t)}\times FT\{s(\omega,t+\tau,T)\}_{\delta(t+\tau)}\rangle=\langle FT\{p(\zeta,\tau,T)\}\rangle,
\end{equation}

where we have used the spectral correlation $p(\zeta,\tau,T)$, the auto-correlation of the spectrum compiled from photon pairs with a separation $\tau$ along the macrotime axis and a separation $T$ from the laser excitation pulse. We note that equation \ref{eqn:five} is only valid for $\delta(t)\approx \delta(t+\tau)$, i.e. for inter-photon lag-times much shorter than the dither period. Explicitly, for a microtime- and macrotime-variant spectrum $s(\omega,t,T)$, the spectral correlation $p(\zeta,\tau,T)$ can be written as

\begin{equation}\label{spec:corr_central}
p(\zeta,\tau,T)=\left\langle\int_{-\infty}^{\infty}s(\omega,t,T)*s(\omega+\zeta,t+\tau,T) d\omega\right\rangle.
\end{equation}

Equation \ref{spec:corr_central} describes the central observable in lifetime-resolved PCFS and can be understood intuitively as a histogram of photon-pairs with a microtime $T$, a macrotime separation of $\tau$, and an energy separation $\zeta$. The form and interpretation of $p(\zeta,\tau,T)$ are dependent on the dynamics of the emissive system and are discussed in the manuscript.

\section*{The lifetime-distinct spectral correlation}
One instructive system conducive to investigation with lifetime-resolved PCFS is a system with two transition dipoles radiating with distinct oscillator strengths for the states $|A\rangle$ and $|B\rangle$. The different radiative lifetimes result in microtime-dependent relative weights of emission intensity originating from states $|A\rangle$ and $|B\rangle$ under $\delta$-like non-resonant excitation. Substituting $s(\omega,t,T)$ as $s(\omega,t,T)=a(T)s_{A}(\omega,t)+b(T)s_{B}(\omega,t)$, where $a(T)$ and $b(T)$ are the relative probabilities of a given photon originating from either state $|A\rangle$ or $|B\rangle$, and separating into the sum of time-averaged terms $\langle....\rangle$ yields

\begin{equation}\label{spec:corr}
\begin{split}
p(\zeta,\tau,T)=\langle\int_{-\infty}^{\infty}a(T)^{2}s_{A}(\omega,t)*s_{A}(\omega+\zeta,t+\tau) d\omega\rangle\\
+ \langle\int_{-\infty}^{\infty}a(T)b(T)s_{A}(\omega,t)*s_{B}(\omega+\zeta,t+\tau) d\omega\rangle\\
+ \langle\int_{-\infty}^{\infty}b(T)a(T)s_{B}(\omega,t)*s_{A}(\omega+\zeta,t+\tau) d\omega\rangle\\
+ \langle\int_{-\infty}^{\infty}b(T)^{2}s_{B}(\omega,t)*s_{B}(\omega+\zeta,t+\tau) d\omega\rangle,
\end{split}
\end{equation}

for the spectral correlation. We can write more compactly:
\begin{equation}\label{reduced:spec_corr1}
\begin{split}
p(\zeta,\tau,T)=a(T)^{2}p_{AA}(\zeta,\tau)\\
+ a(T)b(T)(p_{AB}(\zeta,\tau)+p_{BA}(\zeta,\tau))\\
+ b(T)^{2}p_{BB}(\zeta,\tau).
\end{split}
\end{equation}

The terms quadratic in $a(T)$ and $b(T)$ represent the spectral auto-correlations of the individual states $p_{AA}$ and $p_{BB}$, while the cross-terms involving $p_{AB}$ represent the cross-correlation of the spectra $s_{A}(\omega,t,T)$ and $s_{B}(\omega,t,T)$. The form of the spectral correlation can be understood intuitively because the spectral correlation is compiled from pairs of photons with origins drawn from the four possible combinations of $|A\rangle$ and $|B\rangle$. Importantly, the left- and right-sided correlations $p_{AB}$ and $p_{BA}$ are not identical unless $s_{A}(\omega)$ and $s_{B}(\omega)$ share the same center frequency $\omega_{0}$ and are symmetric in $\omega$.\\

\section*{A lifetime-distinct static doublet}

We now consider the expected form of the spectral correlation for a \textit{static} doublet, where $s_{A,B}(\omega,t_{1})=s_{A,B}(\omega,t_{2})$ for all $t$, in other words time-invariant spectral lines for each state $|A\rangle$ and $|B\rangle$. Per definition, $C(\tau)=0$ as $\delta\omega(t)=0$ for all $t$. The corresponding terms of the spectral correlation are invariant in $\tau$ and
equation \ref{reduced:spec_corr} reduces to\begin{equation}\label{no_tau:spec_corr}
\begin{split}
p(\zeta,T)=a(T)^{2}p_{AA}(\zeta)\\
+ a(T)b(T)[p_{AB}(\zeta)+p_{BA}(\zeta)]\\
+ b(T)^{2}p_{BB}(\zeta),
\end{split}
\end{equation}
reflecting the absence of any dynamic spectral changes.

We numerically simulate the experiment (see Methods section) to showcase the technique and resulting spectral correlation for the static doublet as shown in Figure \ref{fig:Fig2}. We introduce two states with 100 and 1000 ps radiative lifetimes $T_{1}$ and different coherence times $T_{2}$ of $30$ and $60 \textrm{ps}$ with exponential decoherence (Lorentzian lineshape). The corresponding fluorescence lifetime as compiled from the numerical photon-arrival time data is shown in (a) and reflects the biexponential decay corresponding to the emission from the two states. (b) shows the cross-correlation function $g_{X}^{(2)}(\tau)$ for all photons irrespective of the microtime $T$. The individual correlation functions are $\tau$-invariant as expected in the absence of spectral fluctuations. The degree of anti-correlation increases towards a maximum for $\delta=0$. The spectral correlation $p(\zeta,\tau)$ is shown in the contourplot in (c). The higher noise-level at early $\tau$ in (b) and (c) can be attributed to photon-shot noise. The lower panel in (c) shows the lifetime-resolved PCFS data for $\tau=60\mu \textrm{s}$. The evolution from the narrow (long coherence, fast lifetime) to broad (shorter coherence, slower lifetime) is consistent with the evolution of the emission probabilities from state $|A\rangle$ and $|B\rangle$. We also plot the relative cross-correlation probability of photons originating from $|A\rangle$ and $|B\rangle$. At early $T$, a significant fraction of photon-correlation counts stem from the cross-correlation between $|A\rangle$ and $|B\rangle$ as indicated by the side-peaks in the spectral correlation. These side-peaks are maximal around $T\approx 300  \textrm{ps}$, where the relative weights of emission from states $|A\rangle$ and $|B\rangle$ are equal.\\
In PCFS, the high temporal resolution achieved through photon-correlation comes at the cost of the loss of the absolute phase of the spectral information. In other words, both the asymmetry of the lineshape (not not apply for the systems considered in this work) and the center frequency of $s(\omega)$ is lost in the spectral correlation $p(\zeta)$. The unambiguous reconstruction of $s(\omega)$ from $p(\zeta)$ is therefore impossible and the spectral correlation is typically fit with a model parametrizing a suitable form for the underlying emission spectrum.\citeSI{Cui2013aA,Utzat2019A} Using a similar approach for lifetime-resolved PCFS, the lineshape parameters for the two different states can be extracted by applying a global fit to the spectral correlation of a spectral doublet with relative weights of the emission lines of $a(T)$ and $b(T)$. For the lifetime-distinct doublet, $a(T)$ and $b(T)$ can be extracted as the relative weights of the exponential fits describing the biexponential decay dynamics as shown in Figure \ref{fig:Fig2} (c). We validate this approach by applying a global fit to the slices of the spectral correlation $p(\zeta,\tau=60 \mu \textrm{s}, T)$ along $T$ using equation \ref{equ:four} and assigning a Lorentzian lineshape to $|A\rangle$ and $|B\rangle$ as shown in Figure 2e,f and g.

\section*{Numerical Simulation Methods}

We use a custom MATLAB library to numerically simulate the relaxation-resolved PCFS experiments. All code is made publicly available on github.\citeSI{UtzatGitA}
The numerical simulations have three broad components. \textit{i)} the simulation of the photon-emission of a given system, \textit{ii)} the transcription of these photon-streams into time-tagged single-photon data and \textit{iii)} the PCFS analysis of the time-tagged single-photon data. We first model the photon-streams with the PhotonSimulator.m class by writing the photon-emission stream into arrays containing the macrotime information $t$, microtime information $T$, center wavelength $\lambda$ and a lineshape function $g(t)$.\\
For any given time increment of the experiment $1/f_{rep}$, where $f_{rep}$ is the laser repetition frequency, the probability of photon detection is drawn from a Poissonian distribution for reasonable total average detection count rates of around $10^{5}$ photons/second and laser repetition rates of $f_{rep}=10-20MHz$: $p(\lambda,k)=\frac{\lambda^{k}e^{-\lambda}}{k!}$ with $k=1$ and $\lambda=1x10^{5}cps/f_{rep}$. By omitting higher-order (multi-photon) detection events after the same laser excitation pulse with $k\geq2$, the photon-stream reflects the quantum behavior of the single emitter. The total time of photon emission simulated at each stage position was chosen to be 30-40 s, commensurate with a total integration time for the experiment of around 1-2 h. For the lifetime-distinct doublet, each photon is randomly assigned a state $|A\rangle$ or $|B\rangle$ and the corresponding center wavelength $\lambda_{A,B}$ and $g(t)$ are chosen respectively. For the coupled dipole system, the population is drawn from the microtime-dependent population distribution that changes owing to the relaxation from state $|A\rangle$ to $|B\rangle$. The microtime $T$ is drawn from an exponential distribution with the respective emission lifetime $T1_{A,B}$ as $p(T)\propto e^{-T/T1_{A,B}}$. A single-mode photon exhibits a Lorentzian lineshape, and the corresponding lineshape function $g(t)$ will present as exponential dephasing in the time domain. Our simulations are limited to Lorentzian lineshapes. The adaptation to non-pure single-photons i.e. photons in superposition states of multiple spectral modes, is straightforward through the choice of a Gaussian form for $g(t)$. We introduce Gaussian spectral diffusion by drawing the probability for jump occurrence from the boolean set $\{0,1\}$ with a time-invariant probability for each laser pulse. For each jump, an energy offset $\delta\omega$ is drawn from a Gaussian distribution and added to the color $\lambda$ of the photon of the respective state.\\
The action of the PCFS experiment is to assign photons of a given color, coherence time, and lineshape a detector output flag. This assignment is performed with the function PhotonAnalyer.m, which draws the exit flag from the boolean set $\in\{0,1\}$ with the respective probabilities $p_{0,1}$ given by:

\begin{equation}
p_{0,1}(\delta(t),\lambda)_{g(x)}=1/2\pm g(\delta(t)/c)\textrm{cos}(\frac{2\pi\delta(t)}{\lambda}).
\end{equation}

The resulting photon arrival-time data is presorted according to the microtime interval of choice and the different microtime-sorted photon-streams are cross-correlated using a MATLAB implementation of the multi-tau algorithm to perform the PCFS analysis.\citeSI{Magatti2001A} The details of the PCFS analysis are described in \textit{Utzat et al.}\citeSI{Utzat2019A}.

\setcounter{figure}{0}    

\begin{figure}[htp]
    \centering
    \includegraphics[width=14cm]{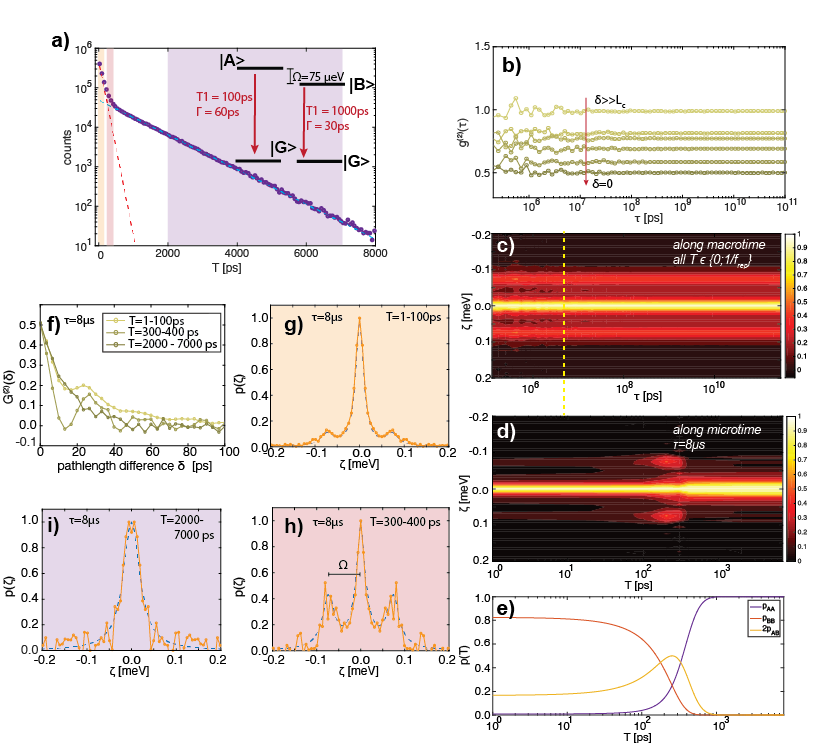}
    \caption{Example system of a static doublet of lifetime-distinct states with different optical coherence times (a, inset). The corresponding photoluminescence lifetime exhibits biexponential behavior (a). The cross-correlation functions $g_{X}^{(2)}(\tau)$ at different stage positions transcribing the degree of spectral coherence into anti-correlations are shown in (b). The invariance of the correlation functions indicates the absence of any spectral fluctuations. The spectral correlation $p(\zeta,\tau)$ is shown in (c) for all $T$. A slice along $T$ for a given $\tau$ is shown in (d). (e) shows the evolution of the weights $p_{AA}$,$p_{BB}$ and $2p_{AB/BA}$ of the different contributions to the spectral correlations, highlighting the possibility of extracting the lineshape parameters from lifetime-resolved PCFS data. The corresponding interferogram is shown in (f) for three different $T$, where $p_{AA}$, $p_{AB,BA}$, and $p_{BB}$ are dominant, respectively. Taking the respective weights into account, the evolution of $p(\zeta,T)$ with $T$ can be fit with two Lorentzian peaks with widths $\Gamma_{1}$, $\Gamma_{2}$ and their energy separation $\Omega$, (g-i).}
    \label{fig:Fig2}
\end{figure}
\newpage

\bibliographystyleSI{ieeetr}
\bibliographySI{SI}

\end{document}